\documentclass[floatfix, noshowpacs, preprintnumbers, twocolumn, amsmath, amssymb, aps,superscriptaddress]{revtex4}

\usepackage{bm}
\usepackage{amsmath}
\usepackage{amssymb}
\usepackage{graphicx}
\usepackage{xcolor}
\usepackage{bbold}
\usepackage{url}
\usepackage[caption=false]{subfig}

\newcommand{\bra}[1]{\left< #1 \right|} 
\newcommand{\ket}[1]{\left| #1 \right>}

\begin{document}
\title{Experimental boson sampling in arbitrary integrated photonic circuits}

\author{Andrea Crespi}
\affiliation{Istituto di Fotonica e Nanotecnologie, Consiglio
Nazionale delle Ricerche (IFN-CNR), Piazza Leonardo da Vinci, 32,
I-20133 Milano, Italy}
\affiliation{Dipartimento di Fisica, Politecnico di Milano, Piazza
Leonardo da Vinci, 32, I-20133 Milano, Italy}
\author{Roberto Osellame}
\email{roberto.osellame@polimi.it}
\affiliation{Istituto di Fotonica e Nanotecnologie, Consiglio
Nazionale delle Ricerche (IFN-CNR), Piazza Leonardo da Vinci, 32,
I-20133 Milano, Italy}
\affiliation{Dipartimento di Fisica, Politecnico di Milano, Piazza
Leonardo da Vinci, 32, I-20133 Milano, Italy}
\author{Roberta Ramponi}
\affiliation{Istituto di Fotonica e Nanotecnologie, Consiglio
Nazionale delle Ricerche (IFN-CNR), Piazza Leonardo da Vinci, 32,
I-20133 Milano, Italy}
\affiliation{Dipartimento di Fisica, Politecnico di Milano, Piazza
Leonardo da Vinci, 32, I-20133 Milano, Italy}
\author{Daniel J. Brod}
\affiliation{Instituto de F\'isica, Universidade Federal Fluminense, Av. Gal. Milton Tavares de Souza s/n, Niter\'oi, RJ, 24210-340, Brazil }
\author{Ernesto F. Galv\~{a}o}
\email{ernesto@if.uff.br}
\affiliation{Instituto de F\'isica, Universidade Federal Fluminense, Av. Gal. Milton Tavares de Souza s/n, Niter\'oi, RJ, 24210-340, Brazil }
\author{Nicol\`o Spagnolo}
\affiliation{Dipartimento di Fisica, Sapienza Universit\`{a} di Roma,
Piazzale Aldo Moro 5, I-00185 Roma, Italy}
\author{Chiara Vitelli}
\affiliation{Center of Life NanoScience @ La Sapienza, Istituto
Italiano di Tecnologia, Viale Regina Elena, 255, I-00185 Roma, Italy}
\affiliation{Dipartimento di Fisica, Sapienza Universit\`{a} di Roma,
Piazzale Aldo Moro 5, I-00185 Roma, Italy}
\author{Enrico Maiorino}
\affiliation{Dipartimento di Fisica, Sapienza Universit\`{a} di Roma,
Piazzale Aldo Moro 5, I-00185 Roma, Italy}
\author{Paolo Mataloni}
\affiliation{Dipartimento di Fisica, Sapienza Universit\`{a} di Roma,
Piazzale Aldo Moro 5, I-00185 Roma, Italy}
\author{Fabio Sciarrino}
\email{fabio.sciarrino@uniroma1.it}
\affiliation{Dipartimento di Fisica, Sapienza Universit\`{a} di Roma,
Piazzale Aldo Moro 5, I-00185 Roma, Italy}

\maketitle

\textbf{Photons naturally solve the BosonSampling problem \cite{Aaronson10}: sample the outputs of a multi-photon experiment in a linear-optical interferometer. This is strongly believed to be hard to do on a classical computer \cite{Aaronson10}, and motivates the development of technologies that enable precise control of multi-photon interference in large interferometers \cite{Poli2009,Peru2010,Sans2012}.  Here we report multi-photon experiments in a 5-mode integrated interferometer. We use novel three-dimensional manufacturing techniques to achieve simultaneous control of 25 independent parameters that describe an arbitrary interferometer. We characterize the chip using one- and two-photon experiments, and confirm the quantum mechanical predictions for three-photon interference. Scaled up versions of this setup are the most promising way to demonstrate the computational capability of quantum systems, and may have applications in high-precision measurements and quantum communication \cite{Obri2009}.}

Large-scale quantum computers hold the promise of solving otherwise intractable computational problems such as factoring \cite{Shor97}. Despite all the experimental effort, this prospect is still far from feasible in all the proposed physical implementations \cite{Ladd2010}. It is thus very important to establish intermediate experimental milestones for the field. One such example is the recent study, by Aaronson and Arkhipov \cite{Aaronson10}, of the computational complexity of simulating linear optical interferometers. It is well known that a linear-optical quantum computer, composed only of passive optical elements (such as beam splitters and phase shifters), becomes universal for quantum computation if adaptive measurements are possible \cite{KLM01,Kok2007}. What was shown in \cite{Aaronson10} is that such a device, even without adaptive measurements, produces an output that is hard to simulate classically. This suggests a feasible experiment to demonstrate the computational capabilities of quantum systems, consisting essentially in observing the multi-photon interference of Fock states in a sufficiently large multimode linear optical interferometer. 

More precisely, in \cite{Aaronson10} it was shown that a linear optical quantum process, consisting of (i) input of photons in a Fock state, (ii) unitary evolution implemented only via beam splitters and phase shifters and (iii) simultaneous photon-counting measurement of all modes, cannot be efficiently simulated classically up to some reasonable complexity assumptions. This became known as the BosonSampling problem \cite{Aaronson10}. Efficient simulation, in this context, is understood as an efficient algorithm run on a classical computer that outputs simulated outcomes with a probability distribution which is close to the one obtained experimentally. At the core of the result is the fact that systems of noninteracting bosons evolve according to permanents of matrices \cite{Troyansky96}. Suppose we have $m$ bosonic modes in the initial Fock state
\begin{equation} \label{state}
\ket{S} = \ket{s_1 s_2 ... s_m} = a^{\dagger s_1}_1 a^{\dagger s_2}_2 ... a^{\dagger s_m}_m \ket{\odot},
\end{equation}
where $s_i$ and $a^{\dagger}_i$ denote, respectively, the occupation number and creation operator for mode $i$, and $\ket{\odot}$ denotes the vacuum state. A linear optical evolution can be described by a $m \times m$ unitary transformation $U$ on the space of creation operators, which induces a unitary transformation $U_F$ on the (exponentially larger) Fock space. The probability amplitude associated with input $\ket{S}$ and output $\ket{T} = \ket{t_1 t_2 ... t_m}$ is given by
\begin{equation} \label{permanent}
\bra{T} U_F \ket{S} = \frac{\textrm{per}(U_{S,T})}{\sqrt{s_1! .. s_m! t_1! .. t_m!}},
\end{equation}
where $U_{S,T}$ is the matrix obtained by repeating $s_i$ times the $i^{th}$ column of $U$, and $t_j$ times its $j^{th}$ row \cite{Scheel04}, and per$(A)$ denotes the permanent of matrix $A$ \cite{Valiant79}, which is defined similarly to the determinant, but without negative signs for odd permutations of matrix elements.

Despite the similarity in the definitions, the permanent and the determinant are surprisingly different with respect to their computational complexity. While the determinant of a $n \times n$ matrix can be calculated in polynomial time, the permanent was proven to be computationally $\#$P-hard \cite{Valiant79}, a class of intractable problems which includes the more well-known NP-hard problems. Despite having their dynamics ruled by permanents, non-interacting bosons cannot be directly used to calculate permanents efficiently, as a typical experiment will have an exponentially large number of outcomes, each predicted by a hard-to-estimate, exponentially small probability associated to a permanent by Eq. (\ref{permanent}).

In \cite{Aaronson10} it was estimated that a system of $\sim 20$ photons in $m \sim 400 $ modes would already take noticeably long to simulate classically. At present, the most promising technology for achieving this regime involves inputting Fock states in multi-mode integrated photonic chips \cite{Poli2008,Poli2009,Matt2009,Smit2009,Peru2010,Sans2010,Cres2011,Sans2012}. Given the theoretical hardness-of-simulation results, experiments of this type serve as a milestone for quantum computation, on the way to more ambitious goals such as efficient integer factoring. 

In this Letter we report the experimental implementation of a small instance of the Aaronson-Arkhipov proposal, using up to three photons interfering in a randomly chosen, 5-mode integrated photonic chip. We have made two important choices which provably make the quantum experiment harder to simulate classically \cite{Aaronson10}: we avoid any structure by choosing a random interferometer, and the interferometer has more modes than the number of input photons. Implementing this arbitrary interferometer also serves as a stringent test of our novel manufacturing techniques, which crucially rely on a three-dimensional interferometer design. This allowed us to verify that non-interacting bosons evolve according to the permanent of matrices of size up to $3 \times 3$.

\begin{figure}[t]
\includegraphics[width=8.5cm]{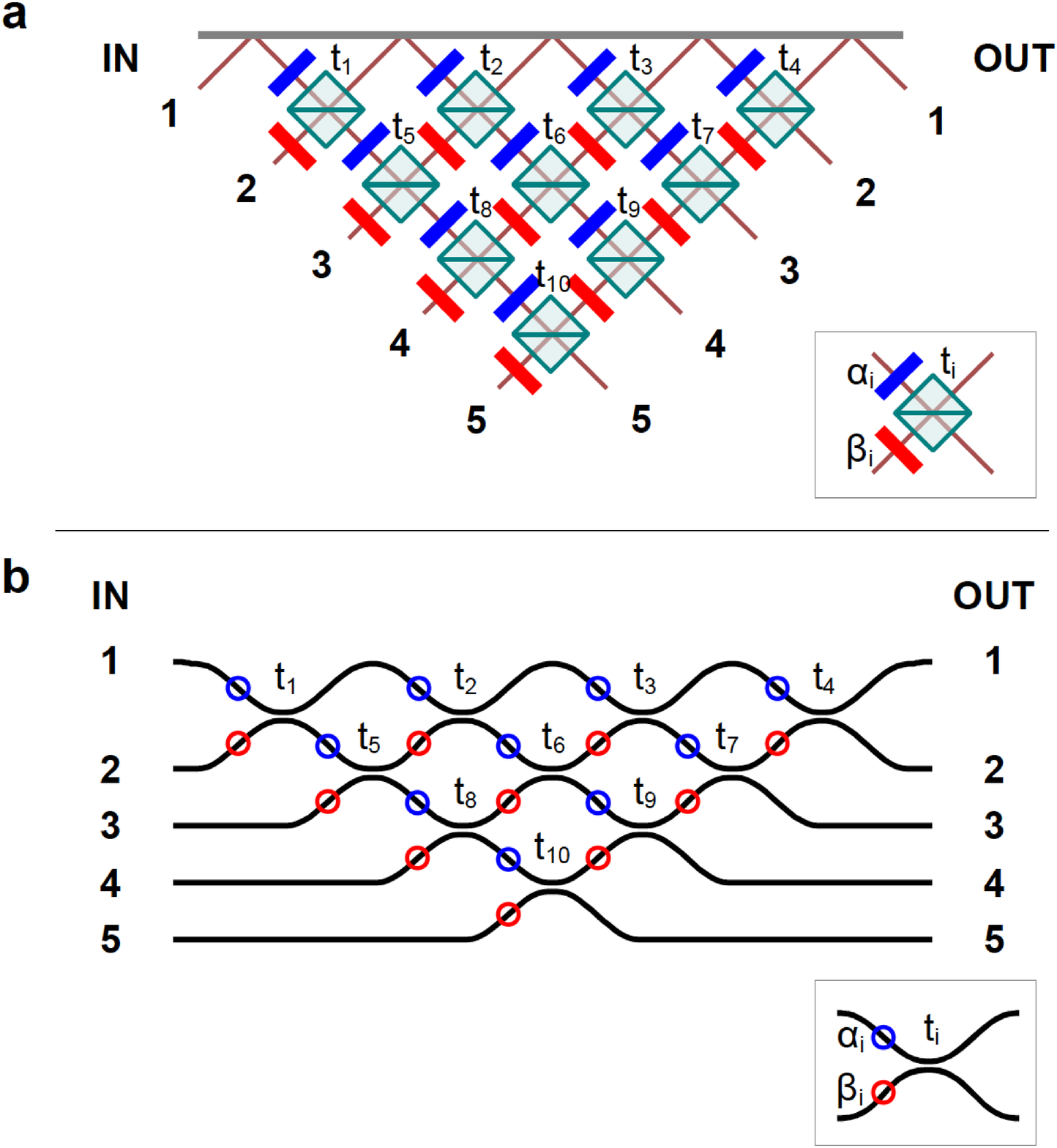}
\caption{\textbf{Layout of multimode interferometers}. \textbf{(a)} Realization of an arbitrary $5 \times 5$ mode transformation via a network of beam splitters with different transmissivities $t_i$. The blue and red boxes stand for different phase shifters. \textbf{(b)} Implementation of the same scheme adopting integrated photonics.
}
\label{layout}
\end{figure}

Any $m$-mode linear interferometer can be decomposed in basic linear optical elements (phase shifters and beam splitters) using the decomposition given in \cite{Reck94}. The general layout of these decompositions is depicted in Fig. \ref{layout}-\textbf{a} for the case $m=5$. It consists of a network of beam splitters with different transmissivities $t_i$ (where $t_i^2$ is the photon's transmission probability), interspersed by phase shifters restricted, without loss of generality, to the $[0,\pi]$ range, as discussed in the Supplementary Information. Unfortunately, building large interferometers out of these discrete elements tends to result in mechanical instabilities which have prevented the demonstration of even a symmetric, 3-mode interferometer that preserves quantum coherence. A more promising approach to obtain stable multi-mode interferometers involves the fabrication of this network of linear optical elements by integrated optical waveguides in a glass chip \cite{Meany12,spa2012}. Waveguides are fabricated using the femtosecond laser micromachining technique \cite{gattass2008flm,dellavalle2009mpd}, which exploits nonlinear absorption of focused femtosecond pulses to induce permanent and localized refractive index increase in transparent materials. Arbitrary three-dimensional circuits can be directly written by translating the sample along the desired path, keeping the velocity constant with respect to the laser beam. This maskless and single-step technique allows fast and cost-effective prototyping of new devices, enabling the implementation of three-dimensional layouts that are impossible to achieve with conventional lithography \cite{Sans2012}. 

\begin{figure*}[ht]
\includegraphics[width=16cm]{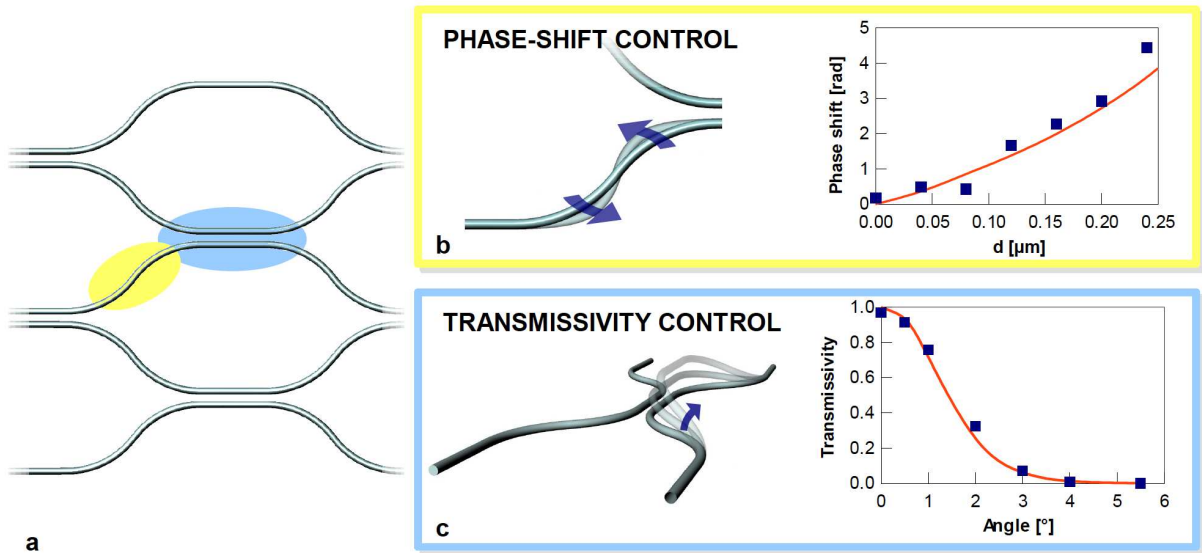}
\caption{\textbf{Independent control of the phase shift and transmissivity at each directional coupler.} \textbf{(a)} Controlled deformation of the S-bent waveguide at the input of each directional coupler and coupling geometry allow independent control over the phase shift and transmissivity. \textbf{(b)} The deformation is given by a non-linear coordinate transformation, which is function of a deformation coefficient $d$. The graph shows the undeformed S-bend together with a deformed one; the experimental dependence of the induced phase shift on the deformation parameter $d$ at $\lambda = 806$ nm is provided, compared to the expected one. \textbf{(c)} Control over the transmissivity of the directional coupler is performed by modulating the coupling coefficient; this is achieved by changing the waveguide spacing in the coupling region by rotating one arm of the directional coupler out of the main circuit plane. A calibration of the transmissivity dependence on the rotation angle at $\lambda = 806$ nm is provided, compared to the theoretical expectation.}
\label{control}
\end{figure*}

In the integrated optics approach the role of beam splitters is performed by directional couplers, devices which bring two waveguides close together to redistribute the light propagating in them by evanescent field coupling \cite{szameit2007,Sans2010}. The integrated optics analogue of the discrete component layout, depicted in Fig. \ref{layout}-\textbf{a}, is shown in Fig. \ref{layout}-\textbf{b}, where one can appreciate the one-to-one correspondence between elements in the two approaches. Our main challenge in implementing the integrated layout of Fig. \ref{layout}-\textbf{b} is to independently control each of the 10 transmissivities $t_i$ and 15 physically relevant $[0,\pi]$ phase shifts $\alpha_i, \beta_i$ of an arbitrary, 5-mode chip. This is because in a typical optical circuit (Fig. \ref{control}-\textbf{a}) changes in the coupler geometry to modulate the transmissivity will change the optical path (and the phase shifts), and vice versa. 

The phase shifters are implemented by deforming the S-bent waveguides at the input of each directional coupler in order to stretch the optical path. The profile of the S-bends is deformed by a suitable coordinate transformation (see Supplementary Information) that stretches the curve in a smooth fashion, to avoid adding waveguide losses, and does not modify the overall footprint of the S-bend, to avoid affecting the transmissivity of the surrounding couplers. Figure \ref{control}-\textbf{b} shows both an undeformed and a deformed S-bend. To determine the amount of phase shift that can be introduced by deforming the S-bend, we fabricated several Mach-Zehnder interferometers with increasingly larger deformations parameterized by the parameter $d$ (see Supplementary information). This allowed us to calibrate our process (Fig. \ref{control}-\textbf{b}) and to verify that a phase shift of up to $\pi$ can be introduced without causing additional losses to the device. 

Achieving an independent control of the transmissivity of each directional coupler is even more difficult. Two parameters that change the transmissivity are the interaction length and the waveguide spacing. Changing either parameter induces, as a side effect, a variation in the optical path that could bring about a significant, and unwanted, phase shift. We overcome this limitation by using our three-dimensional design capability to rotate one arm of the directional coupler out of the main circuit plane, as depicted in Fig. \ref{control}-\textbf{c}. This rotation is an effective way of modifying the waveguide distance in the coupling region (which changes the transmissivity between paths) without affecting the path lengths (and phase shifts). We found that rotation by a few degrees already enables us to span the full range of transmissivity (Fig. \ref{control}-\textbf{c}).   

\begin{figure*}[ht]
\includegraphics[width=0.99\textwidth]{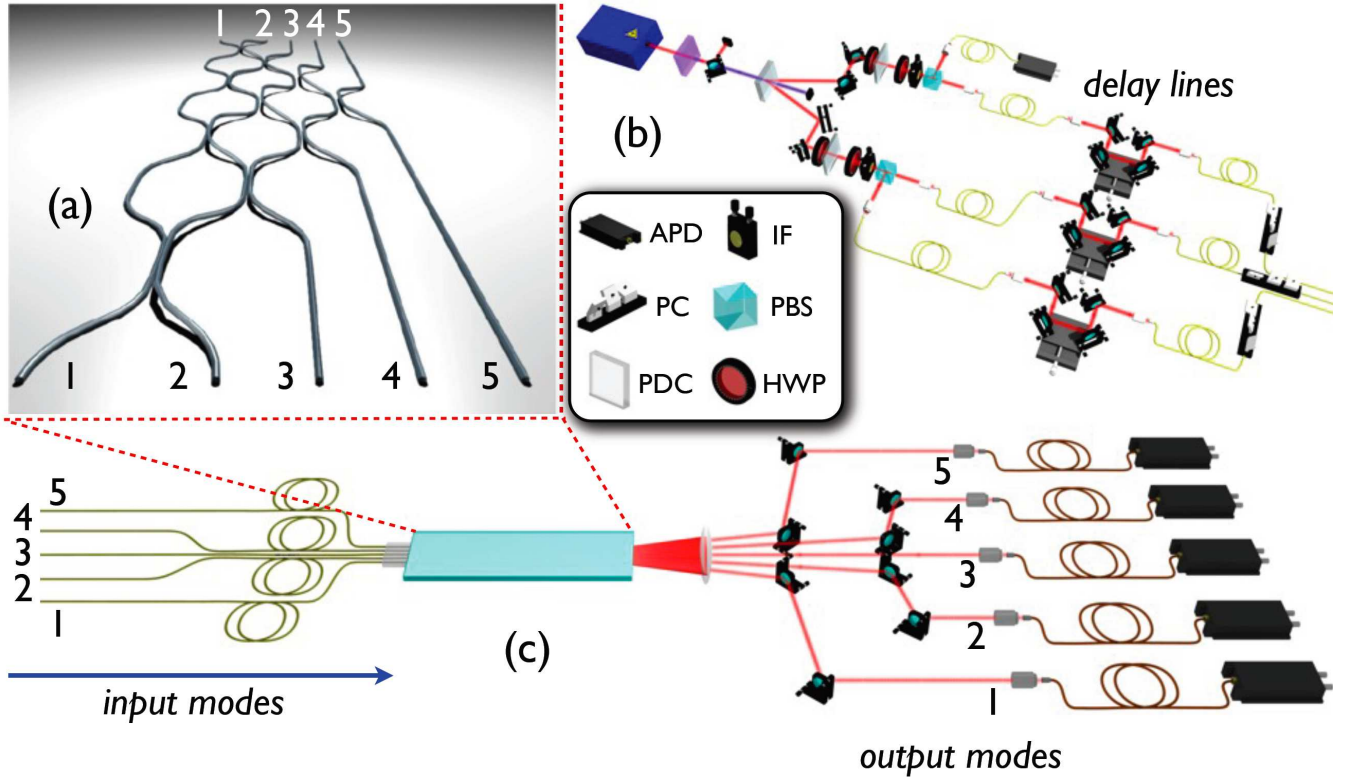}
\caption{\textbf{Experimental setup for the characterization of the chip}. \textbf{(a)} Schematic representation of the interferometer, realized by laser writing technique on a glass substrate. \textbf{(b)} One-photon, two-photon and three-photon states, generated by parametric down-conversion, are injected in the interferometer. The apparatus is composed of the elements: Avalanche Photodiode (APD), Interferential Filter (IF), Polarization Controller (PC), Polarising Beam Splitter (PBS), Parametric Down Conversion (PDC), Half Wave Plate (HWP). Spatial delay lines are adopted to synchronize the three photons. \textbf{(c)} Single-, two- and three-fold coincidence detection at the output ports of the chip is performed to reconstruct the probability of obtaining a given output state realization.}
\label{figura3}
\end{figure*} 

\begin{figure*}[ht]
\includegraphics[width=0.99\textwidth]{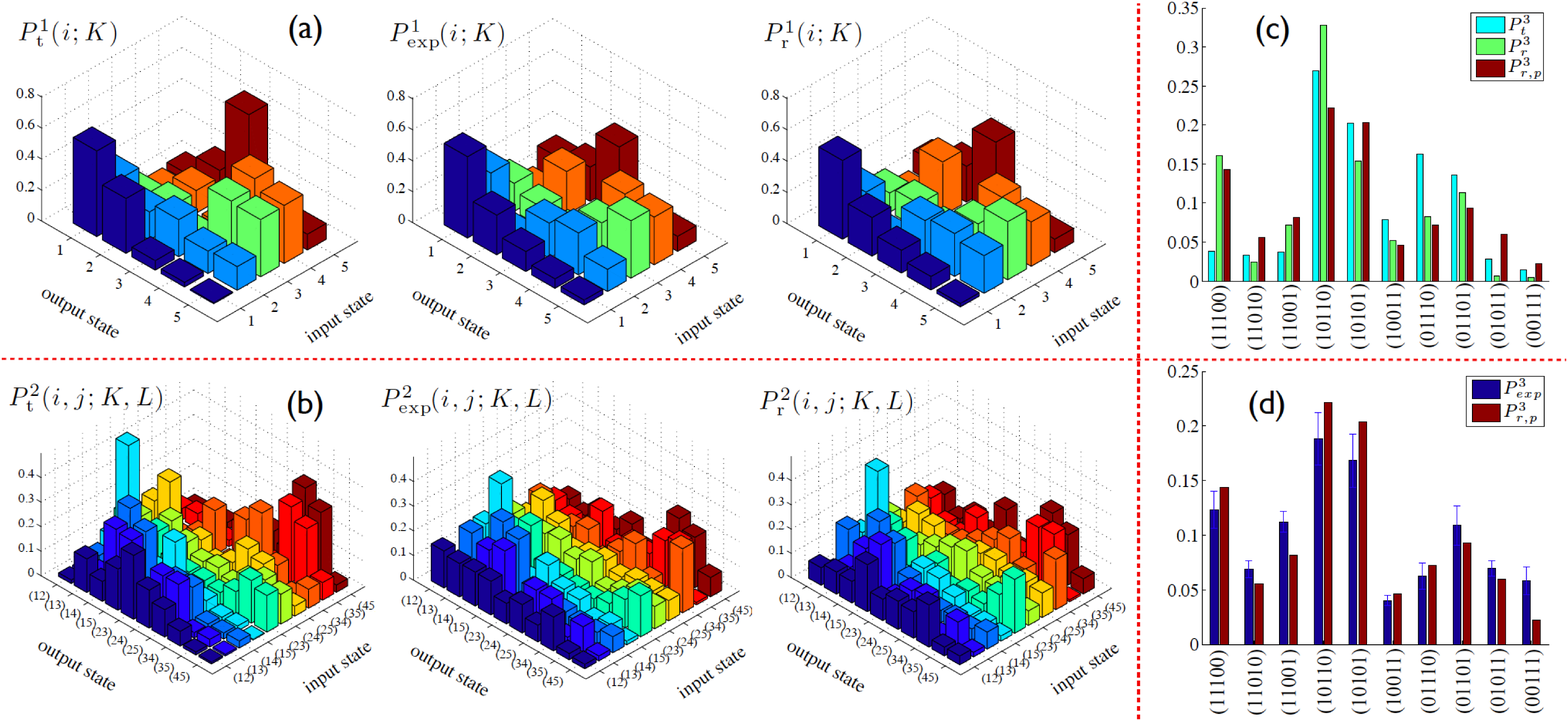}
\caption{\textbf{Experimental results}. \textbf{(a)} One-photon probability distribution: theoretical distribution ($P^1_{t}(i,K)$), experimental distribution ($P^1_{exp}(i,K)$) and reconstructed distribution ($P^1_{r}(i,K)$) . \textbf{(b)}  Two-photon probability distribution: theoretical distribution ($P^2_{t}(i,j,K,L)$), experimental distribution ($P^2_{exp}(i,j,K,L)$) and reconstructed distribution ($P^2_{r}(i,j,K,L)$) .  \textbf{(c)} Expected three-photon probability distribution for an input state $(10101)$: theoretical distribution ($P^3_{t}$), reconstructed distribution ($P^3_{r}$) and reconstructed distribution with partial distinguishability ($P^3_{r,p}$). \textbf{(d)}  Three-photon probability distribution for an input state $(10101)$: experimental distribution ($P^3_{exp}$), reconstructed distribution with partial distinguishability ($P^3_{r,p}$).}
\label{expresults}
\end{figure*} 

To choose which chip to fabricate, we sampled a uniformly random $5 \times 5$ unitary and found its decomposition into directional couplers and phase shifters (see the Supplementary Information for more details). This chip was manufactured and used in single, two-, and three-photon experiments using the sources and detection apparatus described in Fig. \ref{figura3}. As a first step we characterized our 5-mode chip by injecting single photons in each input port $i$ and measuring the probability $P^1_{exp}(i,K)$ of detecting it in output mode $K$. The distribution probability obtained experimentally is shown in Fig. 4--\textbf{a}, together with the theoretical prediction $P^1_{t}(i,K)$ of the sampled unitary $U^t$.  To quantify the agreement between theory and experiment we calculated the similarity between the two distributions, defined for two probability distributions $p, q$ as $S=(\sum_i \sqrt{p_i q_i})^2$. We obtained $S^1_{exp,t}=0.946\pm0.005$, which provides a first confirmation of the proper functioning of the device. Each output probability corresponds to the absolute value of one matrix element of $U^t$.

To obtain a complete characterization of the implemented interferometer we probed the device with pairs of photons. This was done by simultaneously injecting two single photons on all ten combinations of two different input modes $(i,j)$. For each input combination we estimated the ten output probabilities of the photons coming out in two distinct modes $(K,L)$. In all, this corresponds to doing 100 Hong-Ou-Mandel, two-photon interference experiments \cite{HOM87}. The experimental distribution probability $P^2_{exp}(i,j,K,L)$ is reported in Fig. 4-\textbf{b} together with the theoretical distribution $P^2_{t}(i,j,K,L)$ expected from the sampled $U^t$. Each theoretically predicted probability is obtained by calculating the permanent of a $2\times 2$ submatrix of $U^t$. We observe a good agreement between the experimentally obtained probabilities and those given by the permanent formula (2), as evidenced by the similarity $S^2_{exp,t}=0.901\pm0.027$, thus confirming good control over the chip's fabrication parameters.

It is possible to reconstruct the unitary matrix corresponding to the multimode interferometer using only the data corresponding to these one- and two-photon experiments \cite{Obrien12}. We have applied an adapted version of the algorithm described in \cite{Obrien12}, obtaining a reconstructed unitary $U^r$ (see Supplementary Information for details). The similarity between the predictions of our reconstructed $U^r$ and our experimental data is $S^1_{exp,r}=0.990\pm0.005$ (single photon experiments) and $S^2_{exp,r}=0.977\pm0.027$ (two-photon experiments), which indicates a good characterization of the unitary implemented experimentally. 

We have also probed the chip's behavior in the multi-photon regime, by inputting three single photons into modes 1, 3 and 5 of our interferometer, and measuring the probability ratios of all events in which we find photons exiting the chip in three different modes. In Fig.4-\textbf{c} we compare three distributions: the ideal distribution $P^3_t$ obtained from $U^t$; the distribution $P^3_r$ arising from our reconstructed $U^r$ and the one $P^3_{r,p}$ taking into consideration the partial indistinguishability $p$ of the sources we used (for more details refer to the Supplementary Information). Fig.4-\textbf{d} shows a good agreement between the distribution $P^3_{r,p}$ and our experimental results $P^3_{exp}$ as further confirmed by the similarity between these two distributions $S^3_{exp,rp}=0.983\pm0.045$. As these probabilities are proportional to permanents of $3 \times 3$ submatrices of the corresponding unitary, this is an experimental confirmation of the permanent formula (2) in the three-photon, five-mode regime.

We have experimentally confirmed that the permanent formula that governs the quantum mechanical behavior of non-interacting photons holds for up to three photons interfering in a randomly chosen, 5-mode interferometer. Scaling up experiments of this type would provide strong evidence of hard-to-simulate quantum behavior, even in the presence of noise \cite{Rohde12}. This would require developing integrated multi-photon sources \cite{Pate2010}, detectors \cite{Divo2008} and improving the manufacturing process to minimize losses and increase accuracy in the specification of each optical element. The capability of implementing arbitrary unitary transformations may find other applications, such as the discrete Fourier transforms required in the original KLM scheme for linear optical quantum computation \cite{KLM01}, and basis-changing unitaries used in quantum state and process tomography \cite{DurtEBZ10}.

\textbf{Acknowledgements.} This work was supported by FIRB-Futuro in Ricerca HYTEQ and ERC-Starting Grant 3D-QUEST, and by the Brazilian National Institute for Science and Technology of Quantum Information (INCT-IQ/CNPq). We acknowledge support from Giorgio Milani for assessing the data acquisition system.

\textbf{Author contributions.} A.C., R.O., R.R., D.B., E.G., N.S., C.V., P.M., F.S. conceived the experimental approach for hard-to-simulate experiments with integrated photonics. A.C., R.O.,  R.R. developed the technique for 3D circuits, fabricated and characterized the integrated devices by classical optics. N.S., C.V., E.M., P.M., F.S. carried out the quantum experiments. D.B., E.G., N.S., C.V., E.M., F.S. elaborated the data. All the authors discussed the experimental implementation and results and contributed to writing the paper.

\section*{Methods}
\textbf{Femtosecond laser waveguide writing.} The optical circuit for the 5-mode interferometer has been fabricated by direct waveguide writing with the second harmonic ($\lambda =$~515~nm) of a femtosecond Yb:KYW cavity-dumped laser oscillator. Our technique consists in focusing femtosecond laser pulses by a microscope objective (0.6~NA, 50$\times$) in the volume of a transparent borosilicate glass (EAGLE 2000 - Corning). Under suitable irradiation conditions (300~fs pulse duration, 1~MHz repetition rate, 120~nJ energy), this creates a localized refractive index increase that can be exploited to write buried optical waveguides by translating the sample with respect to the laser beam at uniform tangential velocity of 20 mm/s (Aerotech FiberGLIDE 3D air-bearing stages).
Average depth of the fabricated devices under the glass surface is 170~$\mu$m.
The footprint of the 5-mode integrated circuit is 42~mm~$\times$~0.7~mm.

\textbf{Experimental apparatus.} Four photons are produced by parametric down conversion  by pumping a 2mm long BBO crystal by a 392.5nm wavelength pump field \cite{Kwia95}. The four photons are generated at 785 nm, spectrally filtered by 3nm interferential filters and coupled by single mode fibers. One of them acts as the trigger for coincidence detection, while the other three are coupled inside the  chip after passing through different delay lines. The output modes are detected by using multimode fibers and single-photon avalanche photodiodes. Coincidences between different detectors allow us to reconstruct the probability of obtaining a given output state.

\textbf{Reconstructing the chip's unitary.} In order to reconstruct the unitary which best approximates the experimental data, we used an adaptation of the method reported in \cite{Obrien12}. The method in \cite{Obrien12} obtains a unitary that approximates the experimental data, in the form of the full one-photon outcome statistics (sixteen independent parameters, in our 5-mode chip), and a sufficient subset of the two-photon statistics (sixteen additional independent parameters).  By choosing different subsets of two-photon statistics used by the method, we obtained  $25$ different unitaries, each of which best fits the subset of the data used to obtain it. We compared the predictions of these 25 unitaries with the full data, picking the one with best agreement. This served as the starting point for a numerical search to maximize agreement with the experimental data, resulting in the reconstructed unitary $U^r$ reported in the Supplementary Information. 


\begin{thebibliography}{10}

\bibitem{Aaronson10}
S.~Aaronson and A.~Arkhipov.
\newblock The computation complexity of linear optics.
\newblock In {\em Proceedings of the 43rd annual ACM symposium on Theory of
  computing, San Jose, 2011 (ACM press, New York, 2011)}, pages 333--342, 2011.

\bibitem{Poli2009}
A.~Politi, J.~C.~F. Matthews, and J.~L. O'Brien.
\newblock Shor's quantum factoring algorithm on a photonic chip.
\newblock {\em Science}, 325:1221, 2009.

\bibitem{Peru2010}
A.~Peruzzo et~al.
\newblock Quantum walks of correlated photons.
\newblock {\em Science}, 17:1500--1503, 2010.

\bibitem{Sans2012}
L.~Sansoni, F.~Sciarrino, G.~Vallone, P.~Mataloni, A.~Crespi, R.~Ramponi, and
  R.~Osellame.
\newblock Two-particle bosonic-fermionic quantum walk via integrated photonics.
\newblock {\em Phys. Rev. Lett.}, 108:010502, 2012.

\bibitem{Obri2009}
J.~L. O'Brien, A.~Furusawa, and J.~Vuckovic.
\newblock Photonic quantum technologies.
\newblock {\em Nat. Photon.}, 3:687--695, 2009.

\bibitem{Shor97}
P.~W. Shor.
\newblock Polynomial-time algorithms for prime factorization and discrete
  logarithms on a quantum computer.
\newblock {\em SIAM J. Comput.}, 26:1484--1509, 1997.

\bibitem{Ladd2010}
T.~D. et~al. Ladd.
\newblock Quantum computers.
\newblock {\em Nature}, 264:45--53, 2010.

\bibitem{KLM01}
E.~Knill, R.~Laflamme, and G.~J. Milburn.
\newblock A scheme for efficient quantum computation with linear optics.
\newblock {\em Nature}, 409:46--52, 2001.

\bibitem{Kok2007}
P.~Kok, W.~J. Munro, K.~Nemoto, T.~C. Ralph, J.P. Dowling, and G.~J. Milburn.
\newblock Linear optical quantum computing with photonic qubits.
\newblock {\em Rev. Mod. Phys.}, 79:135--174, 2007.

\bibitem{Troyansky96}
L.~Troyansky and N.~Tishby.
\newblock Permanent uncertainty: On the quantum evaluation of the determinant
  and the permanent of a matrix.
\newblock In {\em Proceedings of PhysComp}, 1996.

\bibitem{Scheel04}
S.~Scheel.
\newblock Permanents in linear optical networks.
\newblock {\em arXiv:quant-ph/0406127v1}, 2004.

\bibitem{Valiant79}
L.~G. Valiant.
\newblock The complexity of computing the permanent.
\newblock {\em Theoretical Comput. Sci.}, 8(2):189--201, 1979.

\bibitem{Poli2008}
A.~Politi, M.~J. Cryan, J.~G. Rarity, S.~Yu, and J.~L. O'Brien.
\newblock Silica-on-silicon waveguide quantum circuits.
\newblock {\em Science}, 320:646--649, 2008.

\bibitem{Matt2009}
J.~C.~F. Matthews, A.~Politi, A.~Stefanov, and J.~L. O'Brien.
\newblock Manipulation of multiphoton entanglement in waveguide quantum
  circuits.
\newblock {\em Nat. Photon.}, 3:346--350, 2009.

\bibitem{Smit2009}
B.~J. Smith, D.~Kundys, N.~Thomas-Peter, P.~G.~R. Smith, and I.~A. Walmsley.
\newblock Phase-controlled integrated photonic quantum circuits.
\newblock {\em Opt. Express}, 17:13516--13525, 2009.

\bibitem{Sans2010}
L.~Sansoni, F.~Sciarrino, G.~Vallone, P.~Mataloni, A.~Crespi, R.~Ramponi, and
  R.~Osellame.
\newblock Polarization entangled state measurement on a chip.
\newblock {\em Phys. Rev. Lett.}, 105:200503, 2010.

\bibitem{Cres2011}
A.~Crespi, R.~Ramponi, R.~Osellame, L.~Sansoni, I.~Bongioanni, F.~Sciarrino,
  G.~Vallone, and P.~Mataloni.
\newblock Integrated photonics quantum gates for polarization qubits.
\newblock {\em Nat. Commun.}, 2:566, 2011.

\bibitem{Reck94}
M.~Reck, A.~Zeilinger, H.~J. Bernstein, and Bertani P.
\newblock Experimental realization of any discrete unitary operator.
\newblock {\em Phys. Rev. Lett.}, 73(1):58--61, 1994.

\bibitem{Meany12}
T.~Meany, M.~Delanty, S.~Gross, G.~D. Marshall, M.~J. Steel, and M.~J.
  Withford.
\newblock Non-classical interference in integrated 3d multiports.
\newblock {\em Opt. Express}, 20(24):26895--26905, 2012.

\bibitem{spa2012}
N.~Spagnolo, C.~Vitelli, L.~Aparo, P.~Mataloni, F.~Sciarrino, A.~Crespi,
  R.~Ramponi, and R.~Osellame.
\newblock Three-photon bosonic coalescence in an integrated tritter.
\newblock {\em arXiv preprint arXiv:1210.6935}, 2012.

\bibitem{gattass2008flm}
R.R. Gattass and E.~Mazur.
\newblock {Femtosecond laser micromachining in transparent materials}.
\newblock {\em Nature Photonics}, 2(4):219--225, 2008.

\bibitem{dellavalle2009mpd}
G.~Della~Valle, R.~Osellame, and P.~Laporta.
\newblock {Micromachining of photonic devices by femtosecond laser pulses}.
\newblock {\em Journal of Optics A: Pure and Applied Optics},
  11:013001--013018, 2009.

\bibitem{szameit2007}
A.~Szameit, F.~Dreisow, T.~Pertsch, S.~Nolte, and A.~Tuennermann.
\newblock Control of directional evanescent coupling in fs laser written
  waveguides.
\newblock {\em Opt. Express}, 15:1579--1587, 2007.

\bibitem{HOM87}
C.~K. Hong, Z.~Y. Ou, and L.~Mandel.
\newblock Measurement of subpicosecond time intervals between two photons by
  interference.
\newblock {\em Phys. Rev. Lett.}, 59(18):2044--2046, 1987.

\bibitem{Obrien12}
A.~Laing and J.~L. O'Brien.
\newblock Super-stable tomography of any linear optical device.
\newblock Preprint arXiv:1208.2868v1 [quant-ph], 2012.

\bibitem{Rohde12}
P.~P. Rohde and T.~C. Ralph.
\newblock Error tolerance of the boson-sampling model for linear optics quantum
  computing.
\newblock {\em Phys. Rev. A}, 85:022332, 2012.

\bibitem{Pate2010}
R.B. Patel, A.J. Bennett, I.~Farrer, C.A. Nicoll, D.A. Ritchie, and A.J.
  Shields.
\newblock Two-photon interference of the emission from electrically tunable
  remote quantum dots.
\newblock {\em Nature Photon.}, 4:632--635, 2010.

\bibitem{Divo2008}
A.~Divochiy, F.~Marsili, D.~Bitauld, A.~Gaggero, R.~Leoni, F.~Mattioli,
  A.~Korneev, V.~Seleznev, N.~Kaurova, O.~Minaeva, G.~Gol'tsman, K.~G.
  Lagoudakis, M.~Benkhaoul, F.~L{\'e}vy, and A.~Fiore.
\newblock Superconducting nanowire photon number resolving detector at telecom
  wavelength.
\newblock {\em Nature Photon.}, 2:302--306, 2008.

\bibitem{DurtEBZ10}
T.~Durt, B.-G. Englert, I.~Bengtsson, and K.~{\.Z}yczkowski.
\newblock On mutually unbiased bases.
\newblock {\em Int. J. Quantum Information}, 8:535--640, 2010.

\bibitem{Kwia95}
P.~Kwiat, K.~Mattle, H.~Weinfurter, and A.~Zeilinger.
\newblock New high-intensity source of polarization-entangled photon pairs.
\newblock {\em Phys. Rev. Lett.}, 75:4337--4341, 1995.

\end{thebibliography}

\end{document}


\title{Experimental boson sampling in arbitrary integrated photonic circuits\\
- Supplementary Information -}

\author{Andrea Crespi}
\affiliation{Istituto di Fotonica e Nanotecnologie, Consiglio
Nazionale delle Ricerche (IFN-CNR), Piazza Leonardo da Vinci, 32,
I-20133 Milano, Italy}
\affiliation{Dipartimento di Fisica, Politecnico di Milano, Piazza
Leonardo da Vinci, 32, I-20133 Milano, Italy}
\author{Roberto Osellame}
\email{roberto.osellame@polimi.it}
\affiliation{Istituto di Fotonica e Nanotecnologie, Consiglio
Nazionale delle Ricerche (IFN-CNR), Piazza Leonardo da Vinci, 32,
I-20133 Milano, Italy}
\affiliation{Dipartimento di Fisica, Politecnico di Milano, Piazza
Leonardo da Vinci, 32, I-20133 Milano, Italy}
\author{Roberta Ramponi}
\affiliation{Istituto di Fotonica e Nanotecnologie, Consiglio
Nazionale delle Ricerche (IFN-CNR), Piazza Leonardo da Vinci, 32,
I-20133 Milano, Italy}
\affiliation{Dipartimento di Fisica, Politecnico di Milano, Piazza
Leonardo da Vinci, 32, I-20133 Milano, Italy}
\author{Daniel J. Brod}
\affiliation{Instituto de F\'isica, Universidade Federal Fluminense, Av. Gal. Milton Tavares de Souza s/n, Niter\'oi, RJ, 24210-340, Brazil }
\author{Ernesto F. Galv\~{a}o}
\email{ernesto@if.uff.br}
\affiliation{Instituto de F\'isica, Universidade Federal Fluminense, Av. Gal. Milton Tavares de Souza s/n, Niter\'oi, RJ, 24210-340, Brazil }
\author{Nicol\`o Spagnolo}
\affiliation{Dipartimento di Fisica, Sapienza Universit\`{a} di Roma,
Piazzale Aldo Moro 5, I-00185 Roma, Italy}
\author{Chiara Vitelli}
\affiliation{Center of Life NanoScience @ La Sapienza, Istituto
Italiano di Tecnologia, Viale Regina Elena, 255, I-00185 Roma, Italy}
\affiliation{Dipartimento di Fisica, Sapienza Universit\`{a} di Roma,
Piazzale Aldo Moro 5, I-00185 Roma, Italy}
\author{Enrico Maiorino}
\affiliation{Dipartimento di Fisica, Sapienza Universit\`{a} di Roma,
Piazzale Aldo Moro 5, I-00185 Roma, Italy}
\author{Paolo Mataloni}
\affiliation{Dipartimento di Fisica, Sapienza Universit\`{a} di Roma,
Piazzale Aldo Moro 5, I-00185 Roma, Italy}
\author{Fabio Sciarrino}
\email{fabio.sciarrino@uniroma1.it}
\affiliation{Dipartimento di Fisica, Sapienza Universit\`{a} di Roma,
Piazzale Aldo Moro 5, I-00185 Roma, Italy}

\maketitle

\section{Randomly sampled and reconstructed unitaries}

The matrix $U^{t}$ was sampled from the uniform, Haar distribution over $5 \times 5$ unitary matrices: 
\begin{widetext}
\begin{equation}
U^{t} = \left(
\begin{array}{ccccc}
 0.212 & -0.018+0.165 i & -0.238-0.18 i & -0.429+0.32 i & -0.715+0.2 i \\
 -0.193-0.388 i & -0.045-0.379 i & 0.19 +0.311 i & 0.328 -0.269 i & -0.594+0.03 i \\
 -0.723+0.363 i & 0.087 -0.09 i & -0.076-0.155 i & 0.206 +0.443 i & -0.153-0.193 i \\
 -0.092+0.045 i & -0.148-0.645 i & -0.588+0.184 i & -0.369-0.086 i & 0.167 +0.025 i \\
 0.318 -0.009 i & -0.144-0.594 i & 0.452 -0.405 i & 0.037 +0.387 i & 0.071 +0.025 i
\end{array}
\right),
\end{equation}
\end{widetext}
where the global phase was fixed so as to make $U^{t}_{11}$ real. We then decomposed $U^{t}$ as a product of matrices that act non-trivially on two modes only, each representing a set of one beam splitter and two phase shifters in the range $[0,\pi]$, as described in \cite{Reck94} and depicted in Fig. 1 of the main text. The use of $[0,\pi]$ phase shifters was an adaptation of the decomposition of \cite{Reck94}, which originally used phase shifters in the $[0,2\pi]$ range. This was done to limit the phase shift (and waveguide deformation) introduced by each element, as this may lead to losses. In this new decomposition, only one of the two phase shifters at each beam splitter input branch needs to introduce a non-zero phase shift. Table \ref{Parameters5} reports the parameters obtained in this decomposition.

\begin{table}[ht]\centering \scriptsize
\begin{tabular}{|c||c|c|c|}
\hline 
$i$ & $t_{i}$ & $\alpha_{i}$ [rad] & $\beta_{i}$ [rad] \\
\hline
1 & 0.19 & 0 & 0 \\
\hline
2 & 0.40 & 0.64 & 0 \\
\hline
3 & 0.48 & 0 & 1.37 \\
\hline
4 & 0.44 & 0 & 1.10 \\
\hline
5 & 0.55 & 2.21 & 0 \\
\hline
6 & 0.54 & 0 & 1.02 \\
\hline
7 & 0.51 & 2.93 & 0 \\
\hline
8 & 0.76 & 1.08 & 0 \\
\hline
9 & 0.99 & 2.58 & 0 \\
\hline
10 & 0.95 & 0 & 0\\
\hline
\end{tabular}
\caption{Transmissivities, $t_i$, and phases, $\alpha_i$ and $\beta_i$, related to the layouts in Fig. 1 of the main text, that result from the decomposition of the sampled unitary matrix.}
\label{Parameters5}
\end{table}

Note that all experimental outcomes are invariant under multiplication of $U^{t}$ by a phase shifter at each input and output port. For this reason we have set $\alpha_{1}, \beta_{1}, \beta_{5},\beta_{8}$ and $\beta_{10}$ equal to zero in Table \ref{Parameters5}. 

We have also found a unitary that fits well the single- and two-photon data, in part following the recently proposed method of \cite{Obrien12}. In \cite{Obrien12}, the authors show that the unitary matrix corresponding to a physical interferometer can be reconstructed, up to a round of single phases at the input and output, using only single- and two-photon output probability distributions. The construction requires choosing the matrix elements in one line and one column of the unitary matrix as reference points, and calculating the remaining elements by using a corresponding subset of experimental data consisting of 16 single-photon probabilities and 25 two-photon Hong-Ou-Mandel visibilities. If the physical process is perfectly unitary, this freedom in the choice of the reference elements is irrelevant, as all $25$ possible choices produce the same matrix. 

However, the actual experiment has noise and the data cannot be perfectly described by one unitary matrix. This means that each of the $25$ choices of reference elements, each of which corresponding to using one particular subset of the data, produces a slightly different unitary. For our reconstruction, we have obtained these $25$ unitaries and calculated how well each of them reproduces the whole data set. The measure of this quality was the $\chi^2$ between the whole set of single- and two-photon output predictions for the unitary with respect to the corresponding experimental data. We then chose the matrix which minimized this measure as a starting point for a numerical optimization. This optimization, a combination of brute force and gradient search methods, slightly modified the best unitary obtained by the previous method to fine-tune the agreement with the experimental data. In our reconstruction algorithm, the non-unitary indistinguishability ($q \simeq 0.95$) between the two photons has been taken into account.

Finally, since the reconstruction method of \cite{Obrien12} only obtains the unitary up to a round of arbitrary phases at the input and output modes, we have multiplied each row and column of the optimized unitary so as to obtain the highest gate fidelity with $U^{t}$, so that the matrices are easier to compare to each other. The reconstructed unitary $U^{r}$ is found to be:
\begin{widetext}
\begin{equation}
U^{r} = \left(
\begin{array}{ccccc}
 0.37 & 0.007 +0.151 i & -0.164-0.31 i & -0.442+0.138 i & -0.702+0.099 i \\
 -0.109-0.465 i & -0.013-0.585 i & 0.121 +0.381 i & 0.076 -0.134 i & -0.474-0.147 i \\
 -0.677+0.18 i & 0.134 -0.027 i & -0.283-0.133 i & 0.036 +0.498 i & -0.206-0.319 i \\
 -0.039+0.24 i & -0.08-0.572 i & -0.496-0.046 i & -0.475-0.22 i & 0.265 +0.125 i \\
 0.262 +0.133 i & 0.09 -0.524 i & 0.479 -0.377 i & 0.055 +0.486 i & 0.143 +0.007 i
\end{array}
\right).
\end{equation}
\end{widetext}

Notice that, once again, we have fixed the global phase so as to make $U^r_{1,1}$ real. The gate fidelity with respect to the sampled unitary $U^{r}$ was found to be $F=|Tr(U^tU^{r\dagger})|/5 =0.95$.

\begin{figure}[b]
\includegraphics[width=8.5cm]{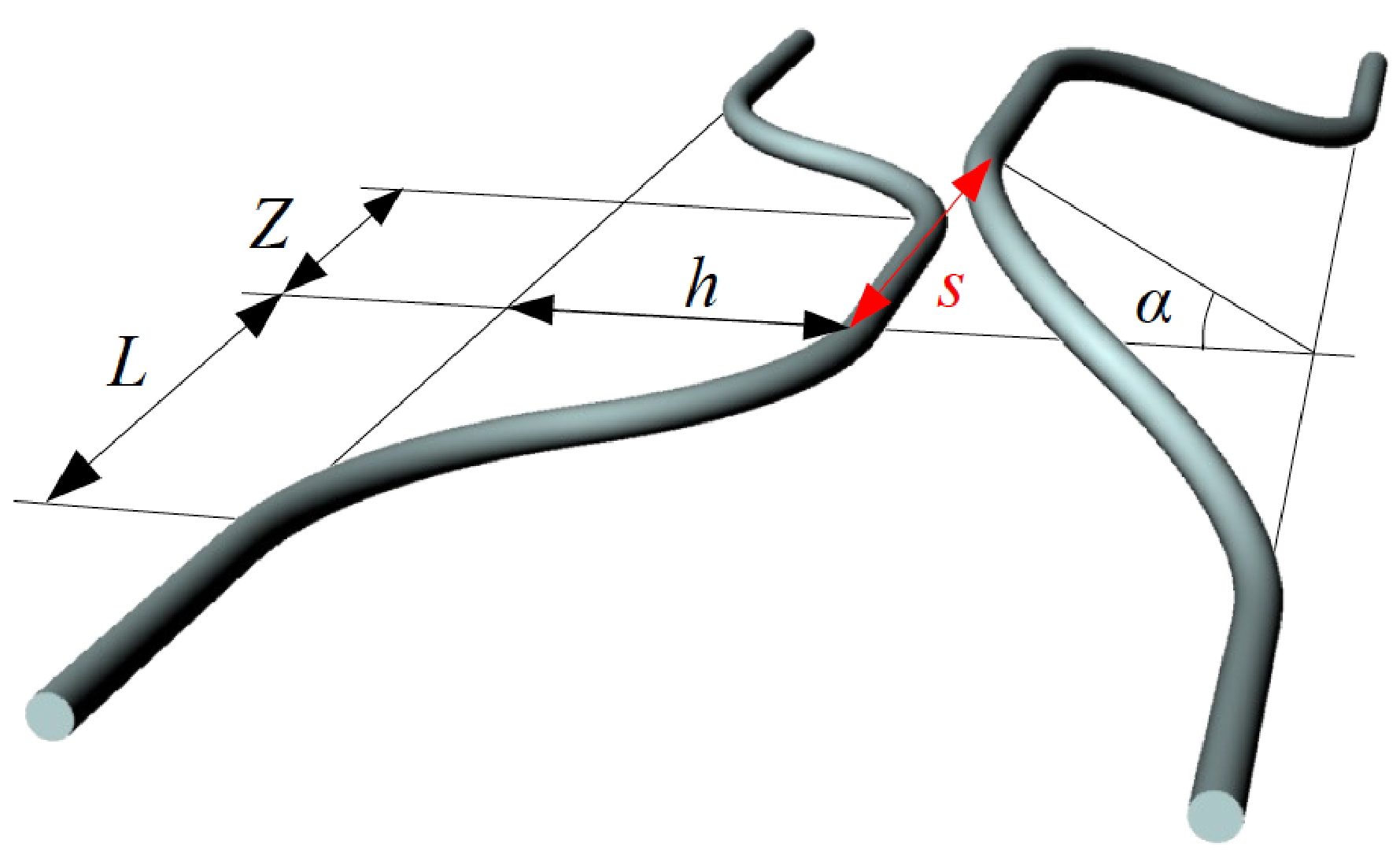}
\caption{\textbf{Definition of the relevant parameters in an interferometer node}. Scheme of a 3D directional coupler constituting the building block of the 5-mode interferometer, where we define all the relevant geometric parameters.
}
\label{Fig1SI}
\end{figure}

\section{Modelling arbitrary phase shifts and transmissivities}
Controlled phase shifts between subsequent directional couplers are implemented by varying the optical path, i.e. by stretching the geometrical length of the connection waveguide. This is perfomed by applying to the S-bent waveguide path the geometrical transformation described in the following.
An undeformed S-bend is described by a sinusoidal function of the kind:
\begin{equation}
y= - \frac{h}{2} \cos \left( \frac{2 \pi}{L} x \right)
\end{equation}
where $h$ and $L$ define the extension of the curve in the two coordinates (Fig. \ref{Fig1SI}). Such curve must be deformed and stretched in a smooth fashion, to avoid adding waveguide losses, and without modifying the overall footprint of the S-bend, which would otherwise affect the position of all the other couplers of the network. Hence, the deformation is operated by the following coordinate transformation:
\begin{equation}
x' = x + d \sin \left( \frac{2 \pi}{L} x \right)
\end{equation}
where $d$ defines the entity of the deformation. The resulting length of the deformed S-bend can be calculated by numerical integration. As discussed and shown in Fig. 2-\textbf{b} of the Main Text, the obtained phase shift was experimentally calibrated with respect to the deformation parameter $d$. A good agreement is observed between the expected phase shift, calculated from the nominal lengthening of the S-bend, and the measured one. The root mean square deviation of the experimentally obtained phase shift with respect to the expected one is 0.25~rad, equivalent to an error in the path-length of about 20~nm.
Independent control on the coupler transmissivity is obtained, on the other hand, by a three-dimensional rotation of one of the coupler's arm. Indeed, this enables to modify the distance between the two waveguides without altering the shape (and length) of the S-bend, which would determine an unwanted phase shift.
To this aim, an accurate modelling of the coupling as a function of the rotation angle is needed.
The transmissivity \footnote{Fraction of power, injected in one waveguide, which is transferred to the other waveguide.} of a directional coupler, can be expressed by \cite{yariv}:
\begin{equation}
T = \sin^2 \left( \kappa Z \right)
\label{eq:T}
\end{equation}
where $\kappa$ is the coupling coefficient and $Z$ is the interaction length (length of the region in which the two waveguides are close to each other) (Fig. \ref{Fig1SI}).
The coefficient $\kappa$ decreases exponentially with the interaction distance $s$ (spacing of the two waveguides within the interaction length) \cite{szameit2007}:
\begin{equation}
\kappa= \kappa_0 e^{-\frac{s}{s_0}}
\label{eq:coupling}
\end{equation}
in which $\kappa_0$ and $s_0$ are suitable constants. These constants have been experimentally measured by fabricating directional couplers in the plane with different $s$ and by fitting the theoretical curve \eqref{eq:coupling} to the experimentally measured transmissions. The values are $\kappa_0=42$ mm$^{-1}$ and $s_0= 2.4~\mu$m. 
On the other hand, the distance between the two waveguides depends on the rotation angle $\alpha$ of the arm of the directional coupler according to:
\begin{equation}
s^2 = h^2 + (h+s_\mathrm{min})^2 - 2 h (h+s_{\mathrm{min}}) \cos \alpha
\label{eq:distance}
\end{equation}
where $h$ is defined above and $s_\mathrm{min}$ is the minimum distance between the waveguides when the two arms are both in the same plane ($\alpha = 0$).
Putting together the Equations \eqref{eq:T},\eqref{eq:coupling},\eqref{eq:distance} one reads:
\begin{equation}
T = \sin^2 \left( e^{\log \kappa_0 L - \frac{1}{s_0} \sqrt{C_1 - C_2 \cos\alpha}}\right)
\label{eq:Tfinal}
\end{equation}
with $C_1= h^2 + \left( h + s_\mathrm{min}\right)^2$ and $C_2 = 2 h \left( h + s_\mathrm{min}\right)$ for brevity.
Thus, inverting the formula \eqref{eq:Tfinal} one can retrieve the angle $\alpha$ needed for achieving a specific value of $T$:
\begin{equation}
\alpha= \arccos \left( \frac{C_1 - s_0^2 \left( \log \kappa_0 Z - \log \arcsin \sqrt{T} \right)^2}{C_2}\right)
\end{equation}
Note that in Fig. 2-\textbf{c} of the Main Text the theoretical curve shown in the graph is exactly Eq. \eqref{eq:Tfinal}, thus displaying the excellent agreement of the experimental points with the model here discussed.

\section{Two-photon and three-photon measurements}

\subsection{Hong-Ou-Mandel visibilities}

The characterization process is based on performing single-photon and two-photon measurements. 
\begin{figure}[ht]
\includegraphics[width=0.45\textwidth]{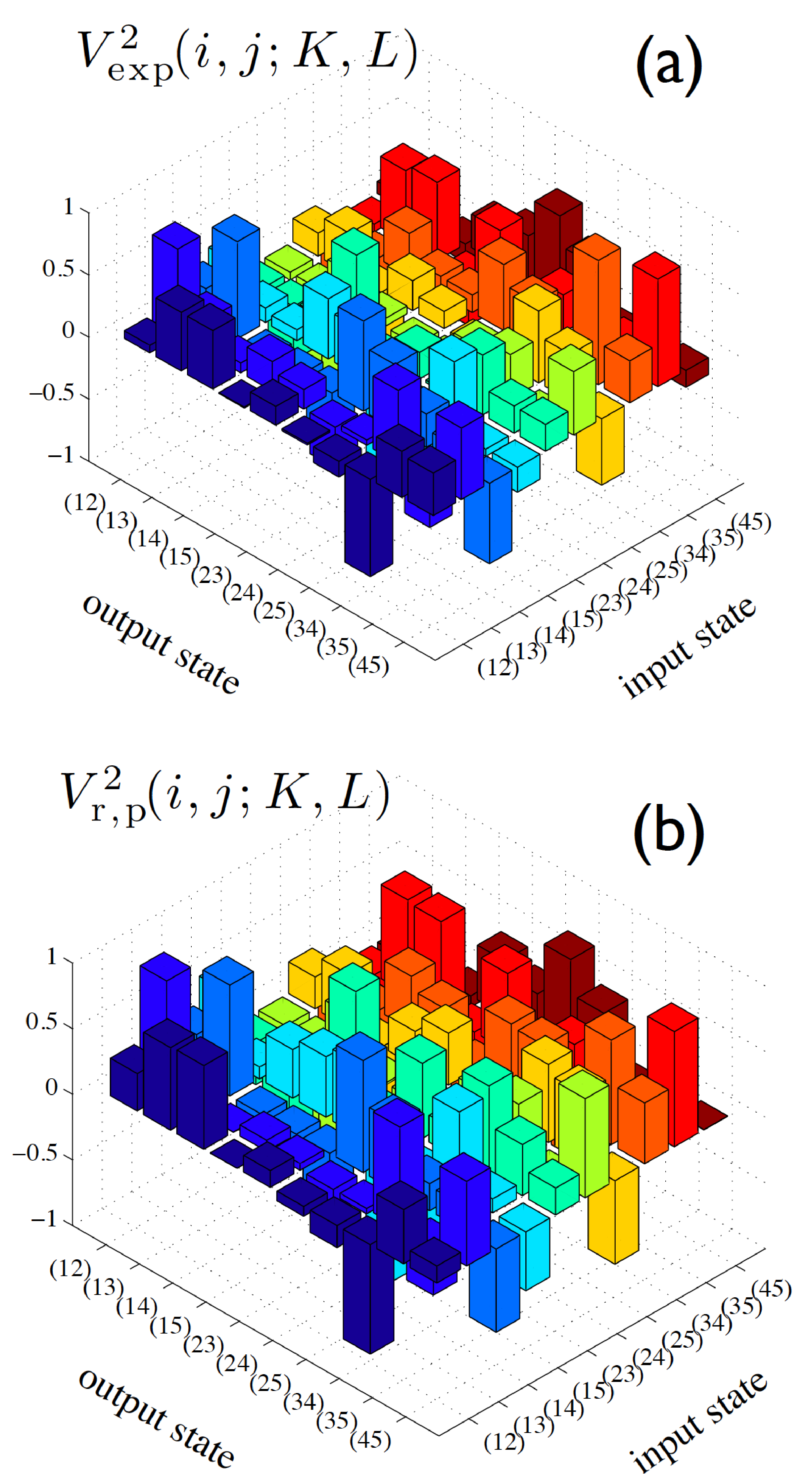}
\caption{\textbf{Two-photon Hong-Ou-Mandel visibilities.} \textbf{(a)} Experimental visibilities $V^{2}_{exp}(i,j;K,L)$ and \textbf{(b)} predictions $V^{2}_{r,p}(i,j;K,L)$ with the reconstructed unitary $U^{r}$ obtained with two-photon measurements.}
\label{fig:visi_exp}
\end{figure} 
The latter consist on injecting all possible combinations of two photons in two different input modes $(i,j)$ and by measuring all possible Hong-Ou-Mandel interferences \cite{HOM87} when the two photons exit in different output ports (K,L). This corresponds to the measurement of a $10\times10$ visibility matrix, where the visibility is defined as:
\begin{equation}
V^{2}(i,j;K,L)=\frac{P^{2,cl}(i,j;K,L)-P^{2}(i,j;K,L)}{P^{2,cl}(i,j;K,L)}.
\end{equation}
Here, $P^{2}(i,j;K,L)$ is the two-photon probability for inputs $(i,j)$ and output $(K,L)$, and $P^{2,cl}(i,j;K,L)$ is the two-photon probability when the two photons are delayed out of the interference region. The complete set of experimentally measured visibilities $V^{2}_{exp}(i,j;K,L)$ is shown in Fig. \ref{fig:visi_exp}-\textbf{a} and is compared with the prediction obtained with the reconstructed unitary matrix $U^{r}$ of the interferometer [Fig. \ref{fig:visi_exp}-\textbf{b}]. The agreement between the two matrices is quantified by the similarity $\mathcal{S}^{2}_{exp,rp}$, which is defined as:
\begin{equation}
\mathcal{S}^{2}_{exp,rp}= 1-\frac{\sum_{i\neq j} \sum_{K \neq L} \vert V^{2}_{exp}(i,j;K,L) - V^{2}_{r,p}(i,j;K,L)\vert }{2 n}
\end{equation}
where $n=100$ is the number of measured visibilities. This parameter achieved a value $\mathcal{S}^{2}_{exp,rp}=0.943 \pm 0.003$ in our experiment, showing the quality of the reconstruction process.

\subsection{Modeling photon distinguishability}

The three-photon state is conditionally prepared by exploiting the second order process of a type-II parametric down-conversion source. The broadband feature of the pump is responsible for a partial distinguishability between the generated photon pairs due to the presence of spectral correlations in the biphoton wave-function. This partial distinguishability introduces a reduction of the interference visibility in the three-photon experiment, and can be modeled by considering an input state of the form:
\begin{equation}
\begin{aligned}
\varrho &= r \vert 1,0,1,0,1 \rangle \langle 1,0,1,0,1 \vert + \\
&+ (1-r) \vert 1_{a},0,1_{b},0,1_{a} \rangle \langle 1_{a},0,1_{b},0,1_{a} \vert.
\end{aligned}
\end{equation}
Here, $r$ is the parameter which takes into account the indistinguishability between the three photons, and the indexes $i=a,b$ label that the photon on input port $3$ belongs to a different photon pair with respect to the photons on input ports $1$ and $5$. The parameter $r$ is proportional to the overlap between the spectral functions of the three photons, and can be written as $r=p^2$, where $p$ is the indistinguishability of two photons belonging to different photon pairs. The latter has been measured directly by performing a Hong-Ou-Mandel two-photon interference in a symmetric $50/50$ beam-splitter, leading to $p=0.63 \pm 0.03$.